\documentstyle[aps,pra]{revtex}
\begin{document}
\draft

\newcommand{\pp}[1]{\phantom{#1}}
\newcommand{\be}{\begin{eqnarray}}
\newcommand{\ee}{\end{eqnarray}}
\newcommand{\ve}{\varepsilon}
\newcommand{\vs}{\varsigma}
\newcommand{\Tr}{{\,\rm Tr\,}}
\newcommand{\pol}{\frac{1}{2}}

\title{
Field quantization by means of a single harmonic oscillator
}
\author{Marek~Czachor}
\address{
Katedra Fizyki Teoretycznej i Metod Matematycznych\\
Politechnika Gda\'{n}ska,
ul. Narutowicza 11/12, 80-952 Gda\'{n}sk, Poland\\
and\\
Arnold Sommerferld Instit\"ut f\"ur Mathematische Physik\\
Technische Universit\"at Clausthal, 38678 Clausthal-Zellerfeld, Germany}
\maketitle

\begin{abstract}
A new scheme of field quantization is proposed. Instead of
associating with different frequencies different oscillators we begin
with a single oscillator that can exist in a superposition of
different frequencies. 
The idea is applied to the electromagnetic radiation field.
Using the standard Dirac-type mode-quantization of the electromagnetic
field we obtain several standard properties such as coherent states
or spontaneous and stimulated emission. As opposed to the
standard approach the vacuum energy is finite and does not have to be
removed by any ad hoc procedure. 
\end{abstract}


\section{Harmonic oscillator in superposition of frequencies}

The standard quantization of a harmonic oscillator is based on 
quantization of $p$ and $q$ but $\omega$ is a parameter. To have,
say, two different frequencies one has to consider two independent 
oscillators. On the other hand, it is evident that there exist
oscillators which are in a {\it superposition\/} of different
frequencies. The example is an oscillator wave packet associated with
distribution of center-of-mass momenta. 

This simple observation raises the question of the role of
superpositions of frequencies for a description of a single harmonic
oscillator. We know that  frequency is typically associated with an
eigenvalue of some Hamiltonian or, which is basically the same, with
boundary conditions. A natural way of incorporating different
frequencies into a single harmonic oscillator is by means of the
{\it frequency operator\/} 
\be
\Omega=\sum_{\omega_k,j_k}\omega_k|\omega_k,j_k\rangle
\langle \omega_k,j_k|
\ee
where all $\omega_k\geq 0$. 
For simplicity we have limited the discussion to the discrete
spectrum but it is useful to include from the outset the possibility
of degeneracies. The corresponding Hamiltonian is defined by
\be
H
&=&
\hbar\Omega\otimes \frac{1}{2}\big(a^{\dag}a+aa^{\dag}\big)
\ee
where $a=\sum_{n=0}^\infty\sqrt{n+1}|n\rangle\langle n+1|$. 
The eigenstates of $H$ are $|\omega_k,j_k,n\rangle$ and satisfy 
\be
H|\omega_k,j_k,n\rangle= \hbar\omega_k\Big(n+\frac{1}{2}\Big)
|\omega_k,j_k,n\rangle.
\ee
The standard case of the oscillator whose frequency is just $\omega$ 
coresponds either to $\Omega=\omega\bbox 1$ or to the subspace
spanned by $|\omega_k,j_k,n\rangle$ with fixed $\omega_k=\omega$. 
Introducing the operators 
\be
a_{\omega_k,j_k}=|\omega_k,j_k\rangle\langle \omega_k,j_k|\otimes a
\ee
we find that 
\be 
H&=&
\frac{1}{2}\sum_{\omega_k,j_k}\hbar\omega_k
\Big(a_{\omega_k,j_k}^{\dag}a_{\omega_k,j_k} 
+
a_{\omega_k,j_k}a_{\omega_k,j_k}^{\dag}\Big).
\ee
The algebra of the oscillator is
\be
{[a_{\omega_k,j_k},a_{\omega_l,j_l}^{\dag}]}&=&
\delta_{\omega_k\omega_l}\delta_{j_kj_l}
|\omega_k,j_k\rangle\langle\omega_k,j_k|\otimes \bbox 1\\
a_{\omega_k,j_k}a_{\omega_l,j_l}&=&
\delta_{\omega_k\omega_l}\delta_{j_kj_l}(a_{\omega_k,j_k})^2\\
a_{\omega_k,j_k}^{\dag}a_{\omega_l,j_l}^{\dag}&=&
\delta_{\omega_k\omega_l}\delta_{j_kj_l}(a_{\omega_k,j_k}^{\dag})^2.
\ee
The dynamics in the Schr\"odinger picture is given by 
\be
i\hbar\partial_t|\Psi\rangle
&=&
H |\Psi\rangle=
\hbar\Omega\otimes \big(a^{\dag}a+\frac{1}{2}\bbox 1\big)
|\Psi\rangle.
\ee
In the Heisenberg picture we obtain the important formula
\be
a_{\omega_k,j_k}(t)
&=&
e^{iHt/\hbar}a_{\omega_k,j_k}e^{-iHt/\hbar}\\
&=&
|\omega_k,j_k\rangle\langle \omega_k,j_k|
\otimes
e^{-i\omega_k t} a
=
e^{-i\omega_k t} a_{\omega_k,j_k}(0).
\label{exp}
\ee
Taking a general state
\be
|\psi\rangle=\sum_{\omega_k,j_k,n}\psi(\omega_k,j_k,n)|\omega_k,j_k\rangle
|n\rangle
\ee
we find that the average energy of the oscillator 
is 
\be
\langle H\rangle=
\langle\psi|H|\psi\rangle
=
\sum_{\omega_k,j_k,n}|\psi(\omega_k,j_k,n)|^2
\hbar\omega_k\Big(n+\frac{1}{2}\Big).
\ee
The average clearly looks as an average energy of an {\it ensemble of
different and independent oscillators\/}. The ground state of the
ensemble, i.e. the one with $\psi(\omega_k,j_k,n>0)=0$ 
has energy 
\be
\langle H\rangle=\frac{1}{2}\sum_{\omega_k,j_k}|\psi(\omega_k,j_k,0)|^2
\hbar\omega_k<\infty.
\ee
The result is not surprising but still quite remarkable if one thinks
of the problem of field quantization. 

The very idea of quantizing the electromagnetic field, as put
forward by Born, Heisenberg, Jordan \cite{BHJ} and Dirac \cite{D},
is based on the observation that the mode decomposition of the
electromagnetic energy is analogous to the energy of an ensemble of
independent harmonic oscillators. In 1925, after the work of
Heisenberg, it was clear what to do: One had to replace each
classical oscillator by a quantum one. But since each oscillator had
a definite frequency, to have an infinite number of different
frequencies one needed an infinite number of oscillators. 
The price one payed for this assumption was the 
infinite energy of the electromagnetic vacuum. 

The infinity is regarded as an ``easy" one since one can get rid of
it by redefining the Hamiltonian and removing the infinite term. 
The result looks correct and many properties typical of a {\it
quantum\/} harmonic oscillator are indeed observed in electromagnetic
field. However, once we remove the infinite term by the procedure of
``normal reordering" the resulting Hamiltonian is no longer {\it
physically\/} equivalent to the one of the harmonic oscillators. For
a single oscillator we can indeed add any finite number and the new
Hamiltonian will describe the same physics. But having two or more
such oscillators we cannot remove the ground state energies by a single
shift of energy: Each oscillator has to be shifted by a different
number and, accordingly, we change the energy differences between the
levels of the global Hamiltonian describing the multi-oscillator
system. And this is not just ``shifting the origin of the energy
scale". Alternatively, one can add up all the ground state corrections
and remove the overall energy shift by a different choice of the origin of
the energy scale. This would have been acceptable if the shift
were {\it finite\/}. Subtraction of infinite terms is in mathematics 
as forbidden as division by zero. (Example:
$1+\infty=2+\infty\Rightarrow 1=2$ is as justified as 
$1\cdot 0=2\cdot 0\Rightarrow 1=2$.) 

The oscillator which can exist in superpositions of different
frequencies is a natural candidate as a starting point for Dirac-type
field quantization. 
We do not need to remove the ground state energy since in the Hilbert
space of physical states the correction is finite. The question we
have to understand is whether one can obtain the well known quantum
properties of the radiation field by this type of quantization.

\section{Field operators: Free Maxwell fields}

The energy and momentum operators of the field are defined in analogy to $H$
from the previous section
\be
H &=& 
\sum_{s,\kappa_\lambda}\hbar \omega_\lambda
|s,\vec \kappa_\lambda\rangle \langle s,\vec \kappa_\lambda|
\otimes \frac{1}{2}\Big(a^{\dag}a+a a^{\dag}\Big)\\
&=& 
\frac{1}{2}\sum_{s,\kappa_\lambda}\hbar \omega_\lambda
\Big(a_{s,\kappa_\lambda}^{\dag}a_{s,\kappa_\lambda} 
+a_{s,\kappa_\lambda} a_{s,\kappa_\lambda}^{\dag}\Big)\\
\vec P &=& 
\sum_{s,\kappa_\lambda}\hbar \vec \kappa_\lambda
|s,\vec \kappa_\lambda\rangle \langle s,\vec \kappa_\lambda|
\otimes \frac{1}{2}\Big(a^{\dag}a+a a^{\dag}\Big)\\
&=& 
\frac{1}{2}\sum_{s,\kappa_\lambda}\hbar \vec \kappa_\lambda
\Big(a_{s,\kappa_\lambda}^{\dag}a_{s,\kappa_\lambda} 
+a_{s,\kappa_\lambda} a_{s,\kappa_\lambda}^{\dag}\Big)
\ee
where $s=\pm 1$ corresponds to circular polarizations. Denote 
$P=(H/c,\vec P)$ and $P\cdot x=Ht-\vec P\cdot \vec x$.
We employ the standard Dirac-type definitions for mode quantization in
volume $V$
\be
\hat{\vec A}(t,\vec x)
&=&
\sum_{s,\kappa_\lambda}
\sqrt{\frac{\hbar}{2\omega_\lambda V}}
\Big(a_{s,\kappa_\lambda}
e^{-i\omega_\lambda t} \vec e_{s,\kappa_\lambda}
e^{i\vec \kappa_\lambda\cdot \vec x}
+
a^{\dag}_{s,\kappa_\lambda}e^{i\omega_\lambda t} 
\vec e^{\,*}_{s,\kappa_\lambda}
e^{-i\vec \kappa_\lambda\cdot \vec x}
\Big)\\
&=&
e^{iP\cdot x/\hbar} \hat{\vec A} e^{-iP\cdot x/\hbar}\\
\hat{\vec E}(t,\vec x)
&=&
i\sum_{s,\kappa_\lambda }
\sqrt{\frac{\hbar\omega_\lambda}{2V}}
\Big(
a_{s,\kappa_\lambda}e^{-i\omega_\lambda t} 
e^{i\vec \kappa_\lambda\cdot \vec x}
\vec e_{s,\kappa_\lambda}
-
a^{\dag}_{s,\kappa_\lambda}e^{i\omega_\lambda t} 
e^{-i\vec \kappa_\lambda\cdot \vec x}
\vec e^{\,*}_{s,\kappa_\lambda}
\Big)\\
&=&
e^{iP\cdot x/\hbar} \hat{\vec E} e^{-iP\cdot x/\hbar}\\
\\
\hat{\vec B}(t,\vec x)
&=&
i\sum_{s,\kappa_\lambda }
\sqrt{\frac{\hbar\omega_\lambda}{2V}}
\vec n_{\kappa_\lambda}
\times 
\Big(a_{s,\kappa_\lambda}e^{-i\omega_\lambda t} 
e^{i\vec \kappa_\lambda\cdot \vec x}
\vec e_{s,\kappa_\lambda}
-
a^{\dag}_{s,\kappa_\lambda}e^{i\omega_\lambda t} 
e^{-i\vec \kappa_\lambda\cdot \vec x}
\vec e^{\,*}_{s,\kappa_\lambda}
\Big)\\
&=&
e^{iP\cdot x/\hbar} \hat{\vec B} e^{-iP\cdot x/\hbar}.
\ee
Now take a state (say, in the Heisenberg picture)
\be
|\Psi\rangle
&=&
\sum_{s,\vec \kappa_\lambda,n}\Psi_{s,\vec \kappa_\lambda,n}
|s,\vec \kappa_\lambda,n\rangle\\
&=&
\sum_{s,\vec \kappa_\lambda}\Phi_{s,\vec \kappa_\lambda}
|s,\vec \kappa_\lambda\rangle|\alpha_{s,\vec \kappa_\lambda}\rangle
\ee
where $|\alpha_{s,\vec \kappa_\lambda}\rangle$ form a family of 
coherent states:
\be
a|\alpha_{s,\vec \kappa_\lambda}\rangle
=
\alpha_{s,\vec \kappa_\lambda}
|\alpha_{s,\vec \kappa_\lambda}\rangle
\ee
The averages of the field operators are 
\be
\langle\Psi|\hat{\vec A}(t,\vec x)|\Psi\rangle
&=&
\sum_{s,\kappa_\lambda }|\Phi_{s,\vec \kappa_\lambda}|^2
\sqrt{\frac{\hbar}{2\omega_\lambda V}}
\Big(
\alpha_{s,\kappa_\lambda}
e^{-i\kappa_\lambda\cdot x} 
\vec e_{s,\kappa_\lambda}
+
\alpha^*_{s,\kappa_\lambda}
e^{i\kappa_\lambda\cdot x}
\vec e^{\,*}_{s,\kappa_{\lambda}}
\Big)\\
\langle\Psi|\hat{\vec E}(t,\vec x)|\Psi\rangle
&=&
\sum_{s,\kappa_\lambda }|\Phi_{s,\vec \kappa_\lambda}|^2
\sqrt{\frac{\hbar\omega_\lambda}{2V}}
\Big(
\alpha_{s,\kappa_\lambda}(0)e^{-i\kappa_\lambda\cdot x}
\vec e_{s,\kappa_\lambda}
-
\alpha^*_{s,\kappa_\lambda}(0)e^{i\kappa_\lambda\cdot x}
\vec e^{\,*}_{s,\kappa_{\lambda}}
\Big)\\
\langle\Psi|\hat{\vec B}(t,\vec x)|\Psi\rangle
&=&
i\sum_{s,\kappa_\lambda }|\Phi_{s,\vec \kappa_\lambda}|^2
\sqrt{\frac{\hbar\omega_\lambda}{2V}}
\Big(
\alpha_{s,\kappa_\lambda}e^{-i\kappa_\lambda\cdot x}
\vec n_{\kappa_\lambda}
\times 
\vec e_{s,\kappa_\lambda}
-
\alpha^*_{s,\kappa_\lambda}e^{i\kappa_\lambda\cdot x}
\vec n_{\kappa_\lambda}
\times 
\vec e^{\,*}_{s,\kappa_{\lambda}}
\Big)
\ee
These are just the classical fields. More precisely, the fields look
like averages 
of monochromatic coherent states with probabilities 
$|\Phi_{s,\vec \kappa_\lambda}|^2$. The energy-momentum operators
satisfy also the standard relations
\be
H
&=&
\frac{1}{2}
\int_V d^3x
\Big(
\hat{\vec E}(t,\vec x)\cdot \hat{\vec E}(t,\vec x)
+
\hat{\vec B}(t,\vec x)\cdot \hat{\vec B}(t,\vec x)
\Big),\\
\vec P
&=&
\int_V d^3x \hat{\vec E}(t,\vec x)\times \hat{\vec B}(t,\vec x).
\ee
It should be stressed, however, that these relations have a
completely different mathematical origin than in the usual formalism
where the integrals are necessary in order to make plane waves
into an orthonormal basis. Here orthogonality follows from the
presence of the projectors in the definition of
$a_{s,\kappa_\lambda}$ and the integration in itself is {\it
trivial\/} since
\be
\hat{\vec E}(t,\vec x)\cdot \hat{\vec E}(t,\vec x)
+
\hat{\vec B}(t,\vec x)\cdot \hat{\vec B}(t,\vec x)
&=&
\hat{\vec E}\cdot \hat{\vec E}
+
\hat{\vec B}\cdot \hat{\vec B}\\
\hat{\vec E}(t,\vec x)\times \hat{\vec B}(t,\vec x)
&=&
\hat{\vec E}\times \hat{\vec B}.
\ee
Therefore the role of the integral is simply to produce the factor
$V$ which cancels with $1/V$ arising from the term $1/\sqrt{V}$
occuring in the mode decomposition of the fields. 
To end this section let us note that 
\be
\langle \Psi|H|\Psi\rangle
&=&
\sum_{s,\kappa_\lambda }
\hbar\omega_\lambda 
|\Phi_{s,\kappa_\lambda}|^2
\Big(
|\alpha_{s,\kappa_\lambda}|^2
+\frac{1}{2}
\Big)\\
\langle \Psi|\vec P|\Psi\rangle
&=&
\sum_{s,\kappa_\lambda }
\hbar\vec \kappa_{\lambda}
|\Phi_{s,\kappa_\lambda}|^2
\Big(
|\alpha_{s,\kappa_\lambda}|^2
+\frac{1}{2}
\Big).
\ee
The contribution from the vacuum fluctuations is nonzero but {\it finite\/}. 

\section{Spontaneous and stimulated emission}

The next test we have to perform is to check the examples that were
responsible for the success of Dirac's quantization in atomic
physics. It is clear that no differences are expected to occur for single-mode
problems such as the Jaynes-Cummings model. 
In what follows we will therefore concentrate on spontaneous and
stimulated emission from two-level atoms.
 
Beginning with the dipole and rotating wave approximations 
we arrive at the Hamiltonian 
\be
H
&=&
\frac{1}{2}\hbar\omega_0\sigma_3
+
\frac{1}{2}\sum_{s,\vec \kappa_\lambda }\hbar\omega_\lambda 
\Big(
a_{s,\vec \kappa_\lambda}^{\dag}a_{s,\vec \kappa_\lambda} 
+a_{s,\vec \kappa_\lambda}a_{s,\vec \kappa_\lambda}^{\dag}\Big)
+
\hbar\omega_0d\sum_{s,\vec \kappa_\lambda }
\Big(
g_{s,\vec \kappa_\lambda}
a_{s,\vec \kappa_\lambda} \sigma_+
+
g_{s,\vec \kappa_\lambda}^*
a_{s,\vec \kappa_\lambda}^{\dag}\sigma_-
\Big)
\ee
where $d\vec u=\langle + |\hat{\vec d}|-\rangle$ is the matrix
element of the dipole
moment evaluated between the excited and ground states, 
and $g_{s,\vec \kappa_\lambda}=i\sqrt{\frac{1}{2\hbar\omega_\lambda V}} 
\vec e_{s,\vec \kappa_\lambda}\cdot \vec u$. The Hamiltonian represents a
two-level atom located at $\vec x_0=0$. 

The Hamiltonian in the interaction picture has the well known form
\be
H_I
&=&
\hbar\omega_0d\sum_{s,\vec \kappa_\lambda }
\Big(
g_{s,\vec \kappa_\lambda}e^{i(\omega_0-\omega_\lambda)t}
a_{s,\vec \kappa_\lambda} \sigma_+
+
g_{s,\vec \kappa_\lambda}^*e^{-i(\omega_0-\omega_\lambda)t}
a_{s,\vec \kappa_\lambda}^{\dag}\sigma_-
\Big).
\ee
Consider the initial state 
\be
|\Psi(0)\rangle
&=&
\sum_{s',\vec \kappa_{\lambda'},m}\Psi_{s',\vec \kappa_{\lambda'},m}
|s',\vec \kappa_{\lambda'},m,+\rangle\nonumber\\
&=&
\sum_{s',\vec \kappa'_{0}}
\Psi_{s',\vec \kappa'_{0},0}
|s',\vec \kappa'_{0},0,+\rangle
+
\sum_{s',\vec \kappa'_{n}}\Psi_{s',\vec \kappa'_{n},n}
|s',\vec \kappa'_{n},n,+\rangle.
\ee
The states corresponding to $n=0$ play a role of a
vacuum. As a 
consequence the vacuum is not represented here by a unique vector,
but rather by a subspace of the Hilbert space of states. It is also
clear that the energy of this vacuum may be nonzero since no normal
ordering of observables is necessary. 

Using the first-order time-dependent perturbative expansion we arrive at
\be
|\Psi(t)\rangle
&=&
|\Psi(0)\rangle
\nonumber\\
&\pp =&
+
\omega_0d
\sum_{s,\vec \kappa_0  }
\frac{e^{-i(\omega_0-\omega_{\lambda_0})t}-1}{\omega_0-\omega_\lambda}
\Psi_{s,\vec \kappa_{\lambda_0},0}
g_{s,\vec \kappa_{\lambda_0}}^*
|s,\vec \kappa_{\lambda_0},1,-\rangle\nonumber\\
&\pp =&
+
\omega_0d
\sum_{s,\vec \kappa_n  }
\frac{e^{-i(\omega_0-\omega_{\lambda_n})t}-1}{\omega_0-\omega_{\lambda_n}}
\Psi_{s,\vec \kappa_{n},n}
\sqrt{n+1}
g_{s,\vec \kappa_n}^*
|s,\vec \kappa_{n},n+1,-\rangle.
\ee
One recognizes here the well known contributions from spontaneous
and stimulated emissions. It should be stressed that although the
final result looks familiar, the mathematical details behind the calculation
are different from what we are accustomed to. For example, instead of
\be
a_{s_1,\vec \kappa_1}^{\dag}|s,\vec \kappa,m\rangle
\sim |s_1,\vec \kappa_1,1;s,\vec \kappa,m\rangle,
\ee
which would hold in the standard formalism for $\vec\kappa_1\neq
\vec\kappa$, we get simply
\be
a_{s_1,\vec \kappa_1}^{\dag}|s,\vec \kappa,m\rangle=0,
\ee
a consequence of 
$a_{s_1,\vec \kappa_1}^{\dag}a_{s,\vec \kappa}^{\dag}=0$. 

\acknowledgements

This work was done mainly during my stay in Arnold Sommerfeld
Institute in Clausthal. I gratefully acknowledge a support from the
Alexander von Humboldt Foundation.

\end{document}